# Impact of Surface Treatment on Noise in PL-Measurements of Silicon Vacancies in 4H-SiC Lateral pin-Diodes

Jannik H. Schwarberg[*,1)], Fabian Magerl[1)], Susanne Beuer[2)], Alexander May[2)], Christian Gobert[2)], Martin Siebert[3)], Christian Miersch[3)], Heino Möller[4)], Wolfgang Knolle[5)], Chihang Luo[6)], Jan F. Dick[1), 2)], Franziska C. Beyer[3)], Mathias Rommel[2)], Jörg Schulze[1), 2)]

1) Chair of Electron Devices, Friedrich-Alexander-Universität Erlangen-Nürnberg, 91058 Erlangen, Germany

2) Fraunhofer Institute for Integrated Systems and Devices Technology, 91058 Erlangen, Germany

3) Department of Energy Materials and Test Devices, Fraunhofer Institute for Integrated Systems and Device Technology, 09599 Freiberg, Germany

4) Intego GmbH, 91058 Erlangen, Germany

5) Leibniz-Institut für Oberflächenmodifizierung, 04318 Leipzig, Germany

6) Department of Modern Physics, University of Science and Technology of China, 230026 Hefei, China

*Email: jannik.schwarberg@fau.de

ABSTRACT – Silicon vacancies ($V_{Si}$) in 4H-SiC are promising candidates for quantum technologies due to their long spin coherence times and integrability into mature semiconductor platforms. However, conventional CMOS-compatible processing introduces significant photoluminescence noise from passivation layers and crystal damage, degrading color center coherence and excitation linewidths. This work evaluates strategies to minimize such background noise. Thermally grown oxides with nitrogen monoxide annealing provide excellent low-noise passivation, remaining stable during subsequent 600 °C thermal treatments. Furthermore, combining reactive ion etching with atomic layer etching eliminates ion-induced surface damage. Into lateral pin-diodes, used for stark shift and photoluminescent excitation linewidth tuning, a selectively etched optical window is integrated. These devices show ideal electrical properties – blocking up to 150 V with leakage current below 10 pA/µm – while significantly enhancing the $V_{Si}$ environment. Single emitters in these pin-diodes show an increased signal-to-noise ratio of 15 for near-surface and of 50 for deeper emitters on both c-plane and a-plane wafers.

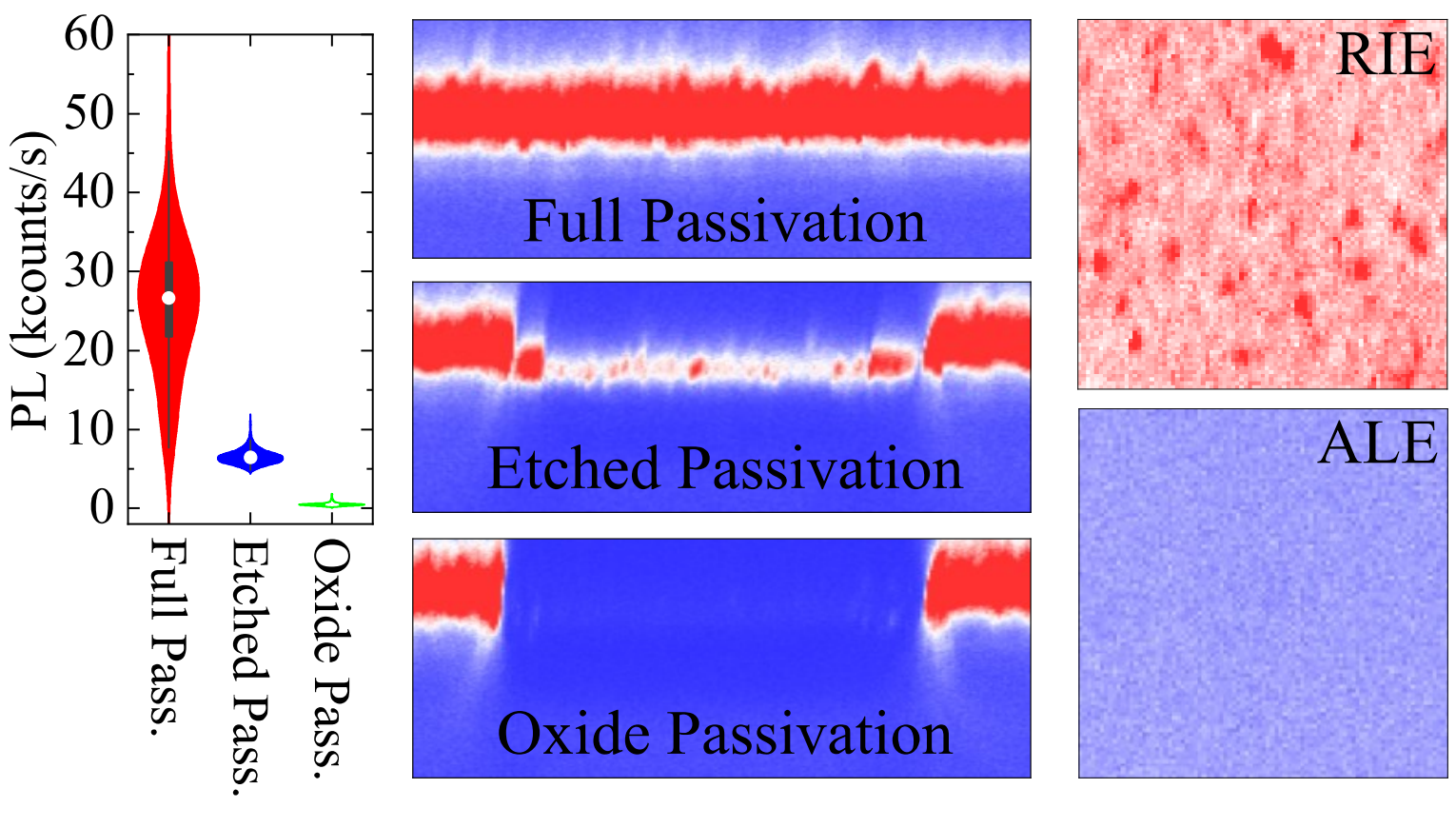

TOC graphic

Silicon vacancies ($V_{Si}$) in 4H-SiC have emerged as promising color centers for applications in quantum sensing, communication, and computing[1–5] due to their favorable spin coherence times and reduced coupling to stray electric fields leading to reduced spectral diffusion[6]. The spin states and their respective photoluminescent excitation (PLE) linewidth can be tuned through integration into semiconductor device structures[2,7]. The advanced and mature processing technology of 4H-SiC[8,9] enables scalable fabrication and the integration of various devices and photonic components on-chip[10–13], making it an attractive platform for applied quantum technologies[1].

Use of $V_{Si}$ for high-precision quantum sensing or computing applications, however, requires a high signal-to-noise ratio (SNR)[14,15]. Since the theoretically possible brightness of $V_{Si}$ is relatively low compared to other color centers[16,17], surface-related luminescence – especially from dry etched or otherwise processed surfaces – often strongly influences the photoluminescence (PL) signal. This effect can outshine the emission especially from near-surface $V_{Si}$. In general, as investigations for nitrogen vacancy (NV) centers in diamond have shown, background signals from the surface degrade the spectral resolution by broadening the resonant PLE linewidth[11–13,18], reducing coherence and relaxation times[19–21] and consequently reduce the overall measurement sensitivity[14,15]. Applying these findings to potential future quantum computation applications of SiC color centers[22] a shorter coherence time also leads to reduced spin manipulation times using quantum gates as well as increased error rates, thus creating a need for more complex error correction finally reducing the number of logical qubits[23].

Integrating SiC color centers into electronic devices like pin-diodes or optical structures from fully integrated quantum photonic circuits is essential for the future of quantum computation and sensing[1]. However, realizing these complex lateral devices requires extensive fabrication flows involving hundreds of processing steps. Processes such as the dry etching of nanophotonic structures[11–13,18,24] or the deposition of insulating[8] and antireflective oxide layers are known to produce optically active defects[25–29]. These defects, forming either in the surface-near crystal or at layer interfaces, generate a strong parasitic PL consisting of a universal background and bright local spots which are shown to originate from a wide range of possibly generated defect color centers. These can be generated either in the SiC itself due to surface near crystal damage stemming from processing or at the interfaces forming due to the deposited or grown insulation layers[26,29].

While this background luminescence may be less critical under low-power resonant excitation, it remains a severe hurdle for the broad range of applications relying on off-resonant excitation. For instance, off-resonant excitation is fundamental to fully integrated quantum sensing chips. Furthermore, it is required for separating excitation from emission in entanglement experiments (e.g., Hong-Ou-Mandel interference[30]) often also utilizing antireflective coatings to reduce measurement times. Lastly, simply locating promising color centers is much easier in low noise environments.

Since the formation of these defect color centers is hard to control and their optical properties are not examined in detail, a promising approach is to try minimizing the generation of these defect color centers and actively generate $V_{Si}$ defects using, for example, electron or ion-irradiation after processing.

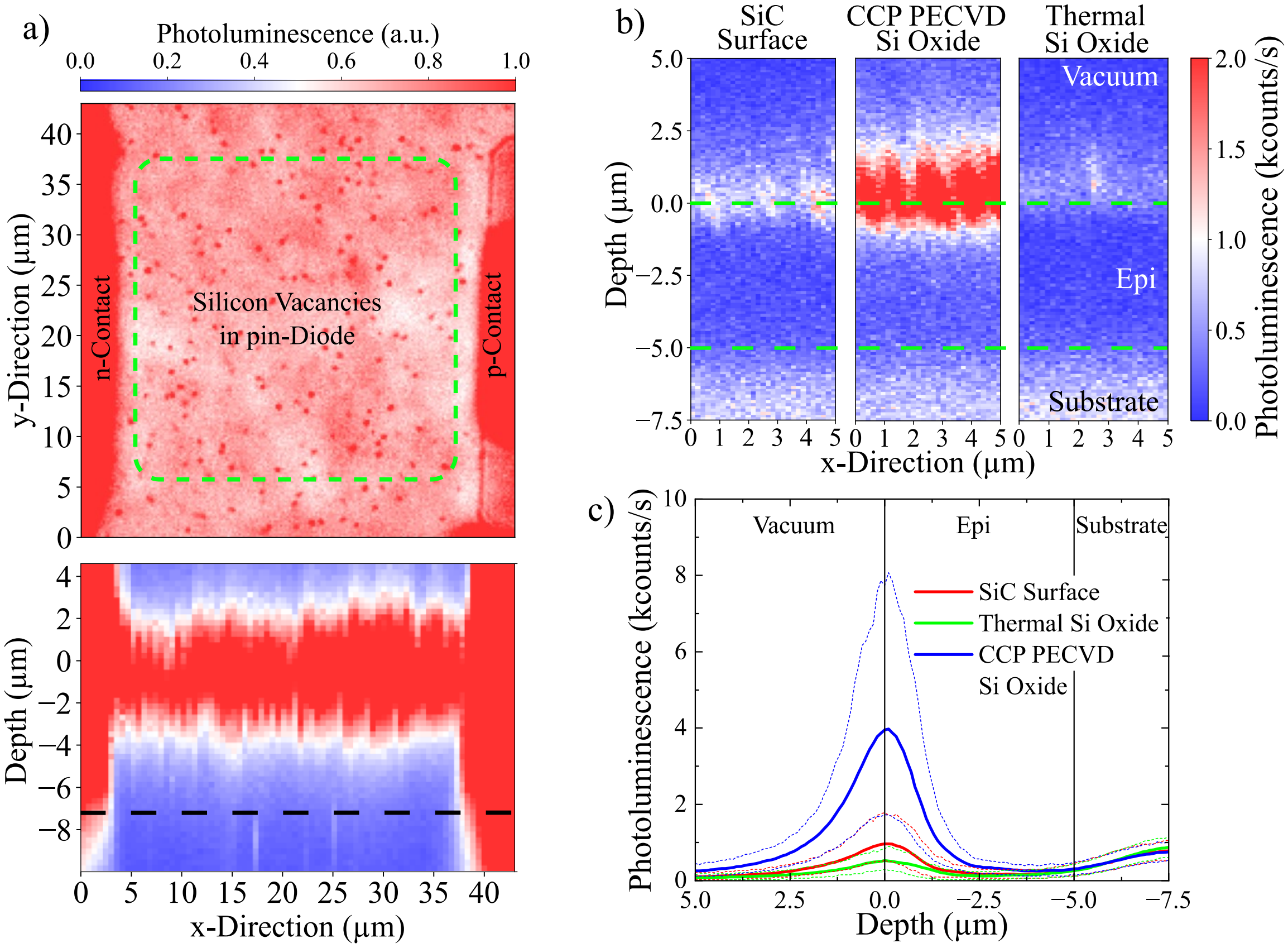

Figure 1. a) PL xy-scan (top) and depth scan (bottom) of a lateral 4H-SiC pin-diode produced in a commercially available CMOS process with generated color centers and high PL-noise. The black dashed line in the depth scan indicates the position of the xy-scan. Note that the graphs have different color coding for better visibility of the color centers in the top graph and the PL gradient over the sample depth in the bottom graph. The color centers in the depth scan appear as long white spots due to limited depth of field. b) Depth scan of three exemplary surfaces showing the difference between a bare SiC surface (left), a surface passivated with either a conductively coupled plasma (CCP) PECVD Si oxide (middle) or a thermally grown Si oxide (right). c) Extracted depth profiles (mean values) from the depth scans with error bands (95% percentile – dashed lines) showing a clear PL peak where the surface/ interface is located.

In Figure 1a a PL scan of a lateral pin-diode manufactured on a 4H-SiC c-plane wafer in a commercially available complementary metal oxide semiconductor (CMOS) process[8] is shown. Color centers were generated using electron irradiation (dose: $1 \cdot 10^{12}$ cm$^{-2}$; energy: 10 MeV with dampening plate leading to effective energy of 4.5 MeV) with subsequent annealing at 600 °C in vacuum for 30 min. The measurement demonstrates very high PL noise originating from the SiC/$SiO_2$ interface and outshining the PL of the only barely observable $V_{Si}$. Especially surface-near color centers were not detectable, as the background PL intensity effectively masks the intrinsic emission of the color centers. This is especially important in nanophotonic structures like waveguides, since they must be very thin to be single mode for the relevant wavelength[12]. Strong background luminance would cover the whole volume of the waveguide making color centers unobservable.

This work focusses on a quantitative comparison of different passivation layers and processing conditions for optimization of PL behavior of the 4H-SiC surface or its interface with the passivation layers. Furthermore, an in-line characterization method for the impact of single process steps on the PL noise is presented. Finally, a possible lateral pin-diode device design using an optical window to reduce PL noise while remaining good electrical properties will be shown. The impact of the reduced noise on the color center quality is estimated using second-order intensity autocorrelation function $g^{(2)}$-measurements.

To investigate the influence of individual processing steps on the measured PL, a series of samples featuring various surface passivation layers and other relevant process treatments were prepared. All samples were fabricated from 5 mm × 5 mm chips of 4H-SiC with a 10 µm thick epitaxial layer (n-type, doping concentration of $3 \times 10^{12}$ cm$^{-3}$)[31], diced from the same region of a single wafer to ensure material uniformity. A complete list of all fabricated samples along with their key processing parameters is provided in Supporting Information section 1. All passivated samples have a comparable passivation layer thickness of roughly 50 nm unless stated otherwise. Note that no specific irradiation for color center generation is carried out. All bright spots observed in the measurements thus can be a wide range of randomly generated defect color centers of different types.

PL measurements were performed using a commercial PL setup built by SQUTEC. A 730 nm laser with 0.25 mW power focused on the sample using a 100x Zeiss objective (NA = 0.9) was used for off-resonant excitation. All measurements (except the recording of the spectrum shown in Figure 5) were carried out at room temperature. The "depth" applied to the y-axis in the depth scans corresponds to the stage movement which does not directly match the focal position of the laser spot inside the sample. Detailed information regarding the PL setup and measurement settings are given in Supporting Information section 2. To minimize the influence of local outliers and ensure reproducibility, three positions with equal sizes distributed across each sample surface were measured. Since the measured PL signal of the surface is highly sensitive to the focal position of the excitation laser, depth-resolved PL profiles were recorded in x-direction as well as y-direction for each measurement spot.

Exemplary depth profiles of a bare 4H-SiC surface, a surface passivated with a plasma-enhanced chemical vapor deposition (PECVD) oxide and a thermally grown oxide with their corresponding depth profiles with error bands are shown in Figure 1b and 1c, respectively. This data illustrates a pronounced impact of surface passivation on the PL noise originating near the surface and for the PECVD oxide spreading deep into the epitaxial layer. In contrast, the PL signal from the underlying epitaxial layer and substrate remains consistent across all samples, confirming the stability and reproducibility of the measurement technique.

Notably, the PL intensity inside the epitaxial layer reaches values as low as 150 counts/s, significantly lower than that of the substrate, indicating the high optical quality of the grown epitaxial layer[31] thus allowing a SNR(in this work defined as the quotient of $V_{Si}$ emission divided by background PL) of up to 50 for $V_{Si}$ (~ 8 kcounts/s). This result holds for both c-plane and a-plane samples (see Supporting Information section 3).

To enable a quantitative comparison across all processed samples, a statistical analysis was performed. For each sample, the distribution of PL intensity values within ±100 nm of the detected surface in all acquired depth scans was extracted. These distributions are visualized as violin plots in Figure 2, providing a clear overview of the surface PL characteristics for each processing condition. The width of the violin plot represents the statistical distribution of PL brightness values for each sample. In these plots, the brightness of the background (bulk of the violin) as well as the various types of defect color centers (height of the violin) can be seen. The focus of discussion in this paper is the background indicated by the bulk of the violin, since the goal is the avoidance of randomly generated defects during processing. The wide variety of defect color centers possibly generated is already subject to other studies[25–29].

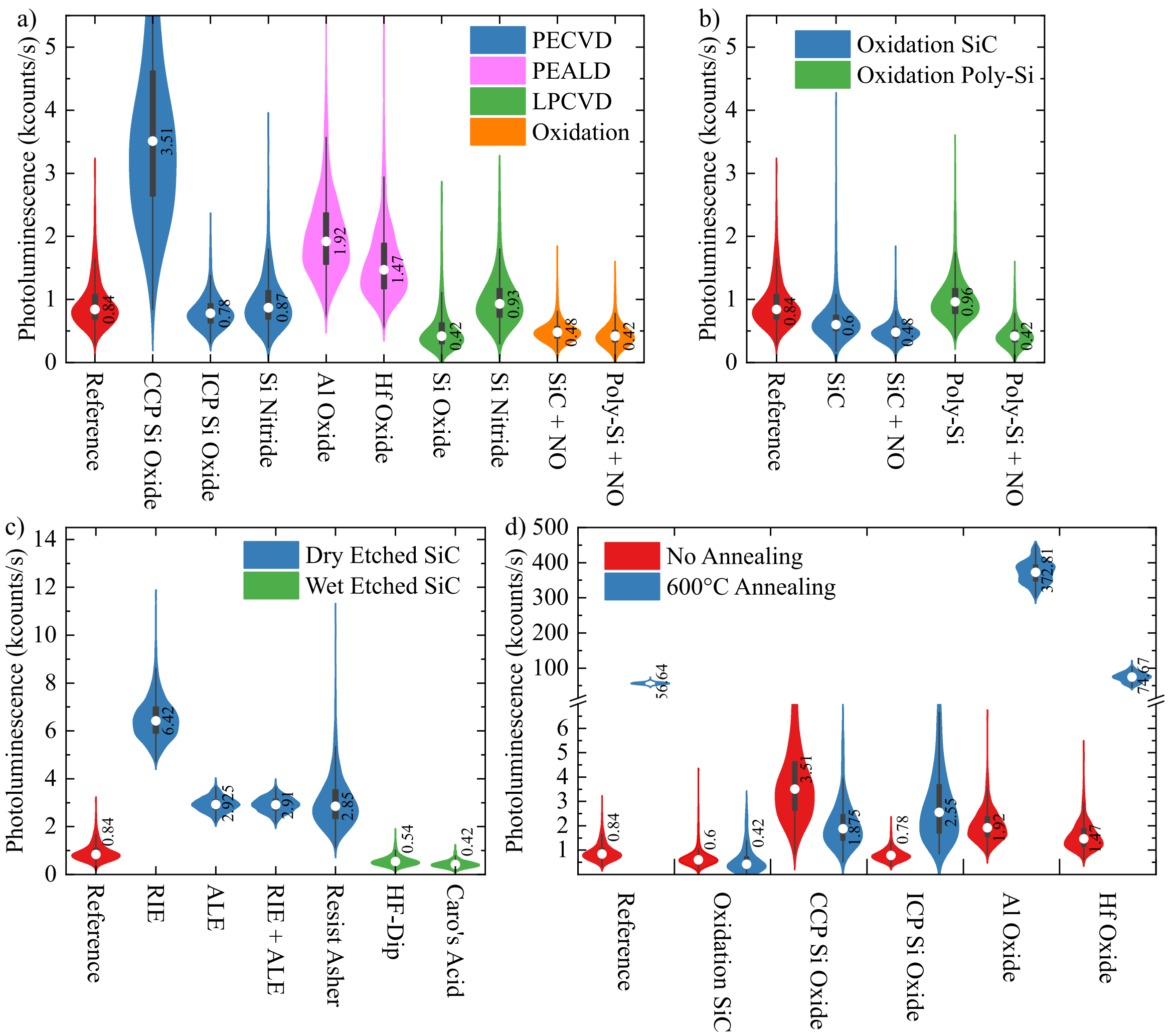

Figure 2. Violin plots of surface or interface PL of all processed samples grouped by a) passivation deposition techniques, b) passivation growth techniques, c) SiC surface modifications and d) influence of post-irradiation anneal at 600 °C in vacuum for 30 min (results before and after annealing are given for each sample). Note the different y-axis scaling for each graph. The median value is included in the graph to facilitate comparison. All abbreviations will be explained in the main text.

Compared to a bare SiC surface with a median background PL of 0.84 kcounts/s, passivation processes involving plasma do not improve but degrade the interface in terms of background PL, as shown in Figure 2a. This enhancement is because plasma-based processes, especially those involving accelerated ions, can induce damage to the SiC surface, resulting in the formation of optically active defects[32].

However, the extent of this effect depends strongly on the specific deposition conditions and the tool employed, as illustrated by the comparison between “CCP Si oxide” and "ICP Si oxide”. The former was deposited using a capacitively coupled plasma (CCP) system, where ions are accelerated perpendicularly toward the substrate, leading to more pronounced interface damage[33] and thus strong background PL of 3.51 kcounts/s (median). In contrast, “ICP Si oxide” as well as the PECVD nitride were deposited using an exclusively inductively coupled plasma (ICP) system, where ions exhibit oscillatory motion and less directional acceleration toward the wafer, resulting in reduced interface damage[33] and, consequently, lower PL background of below 1 kcounts/s.

For the plasma-enhanced atomic layer deposition (PEALD) of aluminum and hafnium oxides, a remote ICP plasma system was employed. In this configuration, plasma is generated remotely from the substrate, minimizing ion bombardment[33]. Therefore, any increase in PL is unlikely to originate from damage inside the SiC itself and is more plausibly attributed to the presence of optically active defects at the interface or within the deposited films themselves. By contrast, thermally activated deposition processes such as low-pressure chemical vapor deposition (LPCVD) do not involve plasma and thus avoid the generation of energetic radicals or ions, finally avoiding generation of optically active defects leading to a low background PL of 0.42 kcounts/s for the oxide layer.

Besides deposition, a passivation can also be grown using thermal oxidation. This process can be performed either directly with the SiC surface[34,35] or utilizing a thin polycrystalline silicon (poly-Si) sacrificial layer deposited on top, which is then oxidized in a self-limiting process at lower temperatures compared to the thermal oxidation of the SiC surface[36]. These processes lead to high quality low noise interfaces with a PL signal of 0.6 kcounts/s for the oxidation of the SiC surface as shown in Figure 2b. This can be further enhanced through post-deposition nitrogen monoxide (NO) annealing, which is known to reduce electrically active interface defects[37,38] and, as demonstrated in Figure 2b, also improves the optical quality of $SiC/SiO_2$ interfaces reducing background PL to below 0.5 kcounts/s for both types.

In addition to surface passivation layers, modifications of the SiC surface itself significantly influence the PL characteristics of the samples. As shown in Figure 2c, a PL measurement conducted shortly (<15 min) after a hydrofluoric acid (HF) dip – which removes the native oxide – reveals a slight decrease in PL intensity. Conducting an HF-Dip with a time-coupled cleaning in Caro's acid to chemically form a thin high quality oxide layer on the surface shows even lower background PL of 0.42 kcounts/s comparable to the thermal oxidation with NO anneal. In contrast to the thermally grown oxide, however, this thin (< 2 nm) and low-density[39] chemical oxide does not protect the surface effectively from degrading in subsequent cleaning or annealing processes, because contaminants can penetrate this thin layer by diffusion and then degrade the interface especially at elevated temperatures.

Plasma-based processes such as dry etching and photoresist ashing (no etching of SiC, only plasma assisted photoresist removal) introduce surface damage, like the effects observed in plasma-enhanced deposition techniques. This damage which increases the background Pl to 6.42 kcounts/s is particularly pronounced in reactive ion etching (RIE), where a DC bias is applied towards the wafer to enable directional etching[40]. Since RIE is commonly used in the fabrication of waveguides[11,12] and optical cavities[18,24], these structures frequently exhibit high PL background noise, which can obscure the characteristic $V_{Si}$ emission and degrade the optical performance of embedded color centers[11–13,18,24].

To mitigate plasma-induced surface damage, we propose the use of atomic layer etching (ALE) for 4H-SiC. ALE offers sub-nanometer etch precision with minimal damage, albeit at low etch rates (~0.2 nm/min). To balance throughput and surface preservation, a two-step process was implemented: an initial RIE etch followed by a 50 nm ALE finishing. The utilization of ALE processes achieves a reduction of background PL intensity to below 3 kcounts/s as illustrated in Figure 2c. Notably a combination of RIE with ALE shows the same surface quality as the ALE without former RIE. Thus, the damage generated with RIE can be diminished completely by ALE. Note that both violins featuring ALE etching results show no neck, indication the absence of defect color centers for this etching technique. A further comparison of these processes is given in Supporting Information section 4).

To enable post-passivation annealing – such as the 600 °C anneal used to improve color center quality[11–13,16,32,41] – it is critical that the passivation layer effectively shields the SiC surface from degradation from chemical reaction with either the annealing environment or residual surface impurities. As shown in Figure 2d, the not passivated reference sample is highly susceptible to optical degradation under these annealing conditions (600 °C, vacuum, 30 min) enhancing the background PL to over 50 kcounts/s which complete outshines the PL of potential $V_{Si}$ and thus makes need for extensive post-anneal cleaning. This also applies to the dry etched sample (not shown). In contrast, samples covered with silicon oxide passivation layers do not show such a strong degradation for identical processing conditions. Especially, the high temperature grown oxides on SiC as well as with the poly-Si layer (not shown) are not influenced.

The as-deposited PEALD aluminum and hafnium oxide films have an amorphous structure[42–44]. The annealing at 600 °C causes the amorphous material to change to a poly-crystalline structure[42–45]. The exact crystallization temperature is reported to be influenced by layer thickness, annealing time and deposition parameters[42]. The crystallization has a huge impact on the optical behavior of these films, as can be observed in Figure 2d. The increased PL of over 372 kcounts/s (Al Oxide) or 74 kcounts/s (Hf Oxide) can either originate from the restructuring of the interface[46] or from defects in the crystalline passivation layer itself, making it a poor choice, at least if post deposition annealing is needed.

Given their favorable optical properties without and with annealing and their low electrically active interface state density and established role in CMOS technology[8,9,38], thermal oxidation of SiC was selected for further process integration and device fabrication.

The complete manufacturing of lateral pin-diodes in 4H-SiC (on both c-plane and a-plane substrates) involves numerous individual steps, each of which carries a risk of degrading the SiC surface's optical quality. To monitor this, a custom-built surface inspection tool (developed by Intego, see Supporting Information section 5) was used to perform in-line PL characterization. The tool raster-scans the full wafer surface under 730 nm excitation and detects emission using an 830 nm long-pass filter. Measurements were conducted before and after critical processing steps, allowing for the evaluation of PL changes induced by each step. The resulting PL maps are presented in Figure 3.

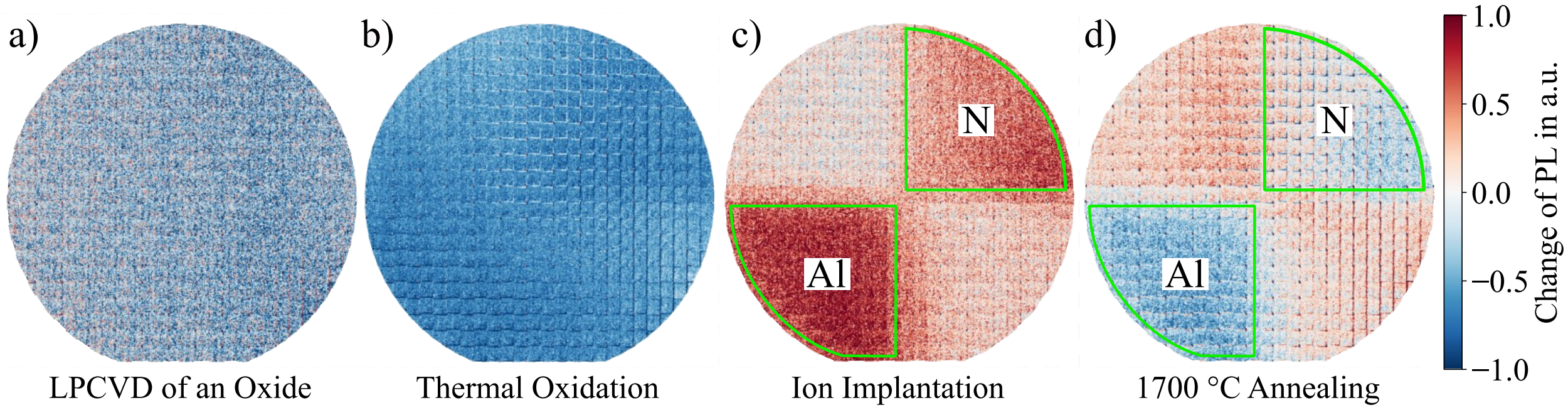

Figure 3. Difference between two wafer scans recorded before and after processing with a custom build surface inspection tool to investigate the change of PL noise. a) LPCVD of $SiO_2$, b) thermal oxidation, c) ion implantation of aluminum and nitrogen in the indicated wafer quarters and d) high temperature annealing at 1700 °C.

The influence of LPCVD of an oxide and thermal oxidation (Figure 3a and 3b) on the PL is consistent with the trends observed in the confocal PL measurements conducted using the SQUTEC setup. As shown in Figure 3c, a significant increase in PL intensity is observed in the ion-implanted regions, attributable to crystal lattice damage caused by ion bombardment[47,48]. This PL enhancement is more pronounced in regions implanted with aluminum, due to the higher atomic mass of aluminum compared to nitrogen, which results in more substantial structural damage during implantation[49]. The same effect can be seen in the confocal PL scan of the pin-diode shown in Figure 5a, where the implanted regions are clearly visible. Subsequent dopant activation annealing at 1700 °C, as illustrated in Figure 3d, partially reverses this effect by healing implantation-induced defects[47,48], thereby reducing the excess PL. These findings demonstrate that in-line PL mapping offers a rapid and non-destructive method for monitoring process-induced changes in the optical properties of the SiC surface. This capability is especially valuable for targeted process optimization and device design refinement, allowing for evaluation of individual fabrication steps and their impact on defect formation and luminescence behavior.

After completing the full CMOS-compatible processing on c-plane as well as a-plane substrates, three different designs of pin-diodes schematically shown in Figure 4a were successfully fabricated. The reference design (top) features the full passivation layer stack, while for the design in the middle and at the bottom these layers are etched down to the SiC surface and the thermally grown oxide, respectively. The details of the material-selective etching process for the bottom design are given in the Supporting Information section 6.

Their corresponding PL-depth scans are shown in Figure 4b. Note that no color centers are generated in these pin-diodes. As can be seen, the optical properties of the different designs are strongly different. The reference design features multiple plasma enhanced deposited layers of different oxides and nitrides. A focused ion beam (FIB) cross-section of the marked area of this design is shown in Figure 4c (top). These diodes show bright PL, even outshining potential color centers in the relevant part of the epitaxial layer of the pin-diode. So, in this design especially color centers close to the surface cannot be detected and used for quantum applications.

For the design where the SiC surface is etched back using RIE, the PL signal is clearly reduced; however, some damage from the directed plasma etching process remains leading to significant surface PL, as discussed in the previous section. In contrast, when a 50 nm thermally grown oxide layer is left for surface passivation, the optical properties remain ideal. Two FIB prepared raster electron microscope (REM) cross-sections at different magnifications are shown in Figure 4c (middle and bottom). These cross-sections confirm successful selective wet chemical etching of the top layers, leaving behind a thermally grown oxide with low roughness. This enables optical characterization with minimal surface noise for all color centers, even those close to the surface.

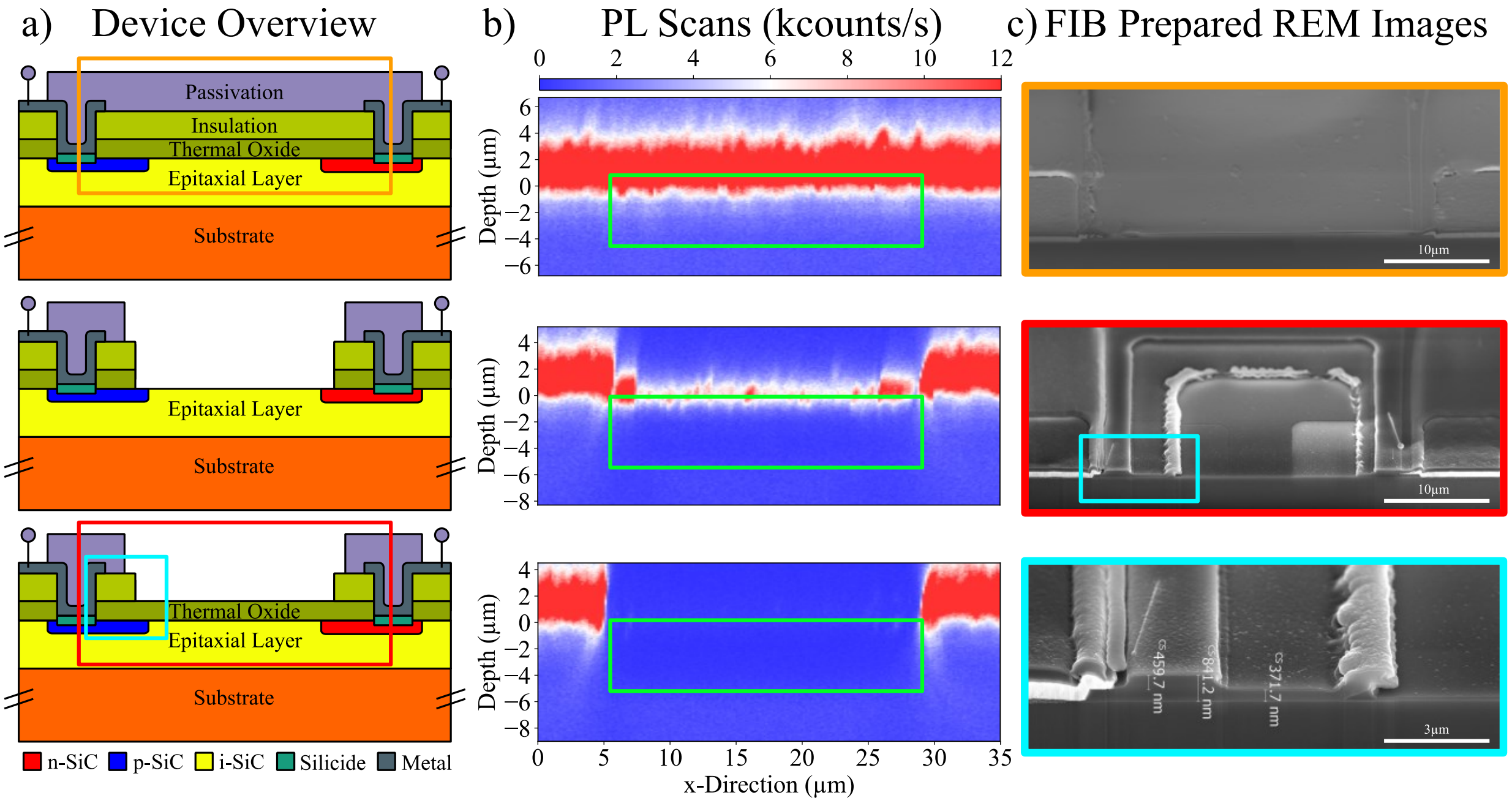

Figure 4. a) Schematic cross-sections of the three different types of processed pin-diodes (top: reference design, middle: passivation layers dry etched down to SiC surface, bottom: passivation layers etched to thermally grown oxide). For details of the etching process see Supporting Information section 6. b) Corresponding PL-depth scans of the pin-diode designs shown in Figure 4a fabricated on 4H-SiC c-plane wafers. The green box highlights the SiC epitaxial layer inside the pin-diode. c) FIB prepared REM cross-sections of the highlighted sections in a). The color coding shows the respective region of the image.

Figure 5a shows a PL scan of an electron-irradiated (dose: $3 \cdot 10^{12}$ cm$^{-2}$; energy: 10 MeV with dampening plate leading to effective energy of 4.5 MeV) lateral pin-diode on a 4H-SiC a-plane wafer, which has the potential to enable resonant excitation across the wafer surface[6] and thus scalable co-integration of further electronic and photonic structures. The xy-scan is recorded exactly at the interface, as indicated by the green dashed line in the depth scan. Near-surface color centers exhibit a SNR of 10–15, while centers deeper in the epitaxial layer reach an SNR of about 40, demonstrating both the excellent optical properties of the passivated surface and the high quality of the epitaxial layer.

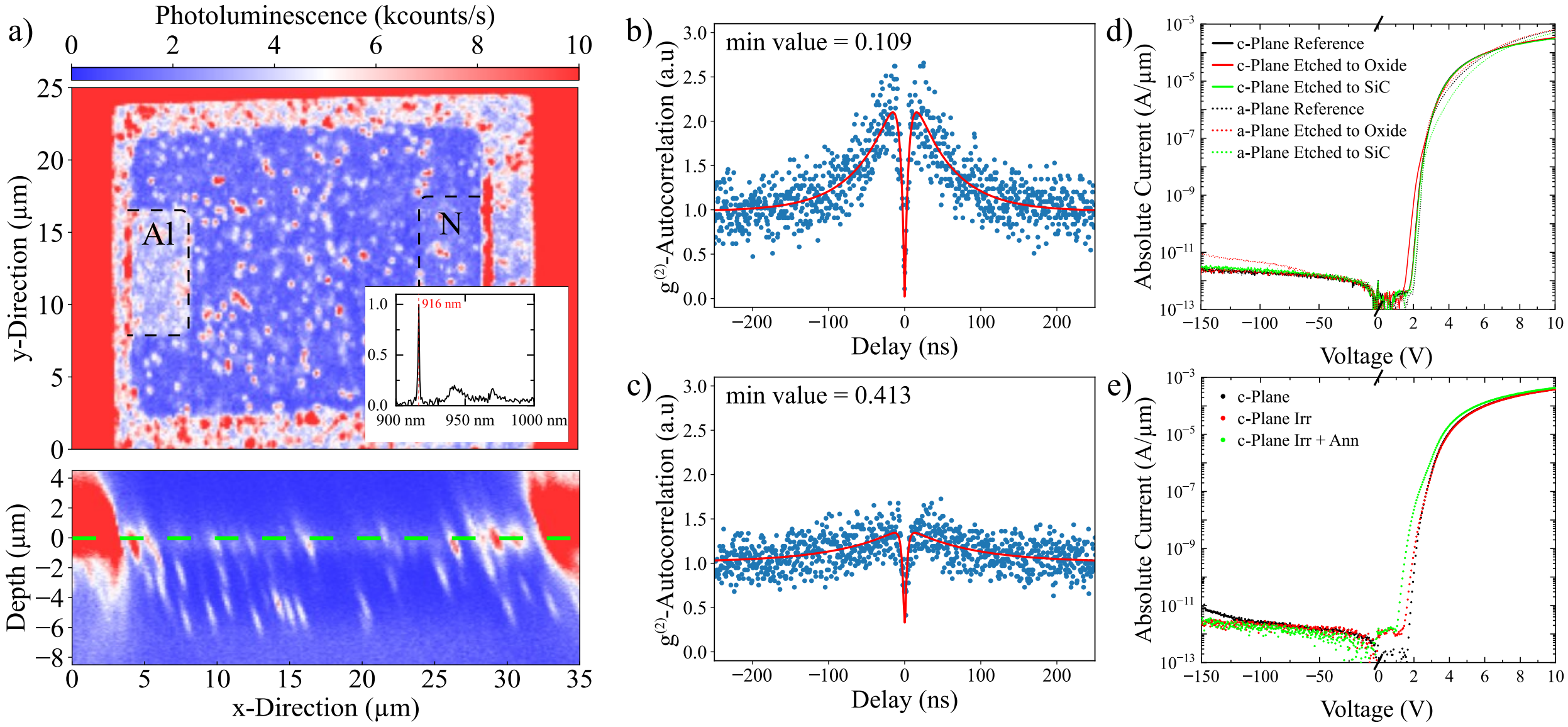

Figure 5. a) PL xy-scan (top) and depth scan (bottom) of an electron irradiated lateral pin-diode on a 4H-SiC a-plane wafer with an optical window etched back to the thermally grown oxide. The slight drift in the depth scan stems from a small hysteresis of the stage. Inset: emission spectrum of selected emitter recorded at 4 K, showing zero-phonon-line at 916 nm, thus confirming successful generation of V2 silicon vacancies. b) and c) show $g^{(2)}$ measurements of $V_{Si}$ inside (top) and outside (bottom) of the optical window below the passivation stack, respectively. d) *I-V* characteristics of the three different discussed pin-diode designs on 4H-SiC c-plane and a-plane wafers. e) Influence of electron irradiation (Irr) and electron irradiation with subsequent annealing at 600 °C in vacuum for 30 min (Irr + Ann) on the *I-V* characteristics of a c-plane pin-diode etched back to the oxide. Note the different x-axis scaling for forward and reverse characteristics.

This finding is supported by the conducted $g^{(2)}$ measurements representatively shown in Figure 5b (inside optical window) and Figure 5c (outside optical window). Data was fit using a three level system[50]. The data is not corrected for background noise, thus the minimum value in the dip indicates the noise level disturbing the measurements. Both measurements show a dip at zero delay below the value of 0.5 thus indicating the presence of a single emitter. Nevertheless, as can be seen, the data for the $V_{Si}$ in the optical window shows the lower minimum value at zero delay. The shoulders can be explained by excitation using a high laser power and thus photon-bunching effects indicating the presence of a metastable state[51,52]. For the $V_{Si}$ outside of the window background PL brings up the dip at zero delay. This effect gets worse the closer to the surface the measurement is carried out up to the point where no dip can be found at all. This low surface noise inside the optical window is one of the requirements for low PLE linewidth[11–13,18], long coherence and relaxation times[19–21] and measurement sensitivity[4,15] finally underlining the importance of high-quality surface treatments.

The electrical properties of the lateral pin-diodes were also examined. The current-voltage (*I-V*) characteristics for different diode designs on c-plane and a-plane wafers (Figure 5d) show only minor substrate- and processing-related differences[53]. All devices exhibit a steep current increase with low ideality factor in forward mode until limited by the series resistance, which is identical for all designs on a given wafer. In blocking mode, the measured currents are mostly influenced by instrument accuracy rather than the device[53]. Consequently, voltages up to 150 V can be applied without significant leakage current, minimizing spin-state disturbance and reducing spectral diffusion, while enabling a large Stark shift of the absorption and emission frequencies[2].

Electron irradiation followed by annealing at 600 °C in vacuum has negligible influence on the electrical performance of lateral pin-diodes with an optical window etched back to the thermally grown oxide, as shown in Figure 5e. Although irradiation and annealing introduce some crystal defects (including color centers and unwanted defects), this only slightly increases current in the forward recombination regime. Blocking characteristics remain unaffected, with currents below 10 pA/µm up to 150 V.

In conclusion, the impact of processing steps and passivation layers on device performance has been investigated. To optimize optical properties, plasma processes should be avoided in optically relevant regions. RIE etching should be followed by ALE etching to minimize surface damage. Both thermally grown oxides (either on SiC or with oxidizing poly-Si thin films) exhibit negligible background PL even after annealing at 600 °C in vacuum for 30 min. Using a thermally grown oxide passivation layer for lateral pin-diodes on a-plane wafers and etching an optical window enables PL characterization with high SNR (up to 15 for near-surface $V_{Si}$) while maintaining excellent electrical properties. Electron irradiation and annealing have negligible effects on *I-V* behavior, allowing strong Stark shifts with minimal electronic noise from leakage currents.

ASSOCIATED CONTENT

Supporting information about the measurement setups used for characterization, experimental details and additional results (PDF)

AUTHOR INFORMATION

**Jannik H. Schwarberg:** Conceptualization, Methodology, Investigation, Validation, Formal analysis, Visualization, Writing – Original Draft; **Fabian Magerl:** Conceptualization, Methodology, Writing – Review & Editing; **Susanne Beuer:** Investigation; **Alexander May:** Investigation; **Christian Gobert:** Investigation; **Martin Siebert:** Investigation; **Christian Miersch:** Investigation, **Heino Möller:** Resources; **Wolfgang Knolle:** Investigation; **Chihang Luo:** Investigation; **Jan F. Dick:** Visualization, Writing – Review & Editing; **Franziska Beyer:** Supervision, Funding Acquisition, **Mathias Rommel:** Supervision, Funding Acquisition, Writing – Review & Editing; **Jörg Schulze:** Supervision, Funding Acquisition

ACKNOWLEDGMENT

We would like to thank Aixtron and Birgit Kallinger for carrying out the epitaxy processes. Special thanks go to the cleanroom teams at LEB and THM for their support in the fabrication of the samples. We are also grateful to Leander Baier for his assistance with the electrical measurements. Finally, we sincerely thank Zongwei Xu and Pengfei Wang for their hospitality and support during the measurements conducted in China.

REFERENCES

1. Moody, G.; Sorger, V. J.; Blumenthal, D. J.; Juodawlkis, P. W.; Loh, W.; Sorace-Agaskar, C.; Jones, A. E.; Balram, K. C.; Matthews, J. C. F.; Laing, A.; Davanco, M.; Chang, L.; Bowers, J. E.; Quack, N.; Galland, C.; Aharonovich, I.; Wolff, M. A.; Schuck, C.; Sinclair, N.; Lončar, M.;

Komljenovic, T.; Weld, D.; Mookherjea, S.; Buckley, S.; Radulaski, M.; Reitzenstein, S.; Pingault, B.; Machielse, B.; Mukhopadhyay, D.; Akimov, A.; Zheltikov, A.; Agarwal, G. S.; Srinivasan, K.; Lu, J.; Tang, H. X.; Jiang, W.; McKenna, T. P.; Safavi-Naeini, A. H.; Steinhauer, S.; Elshaari, A. W.; Zwiller, V.; Davids, P. S.; Martinez, N.; Gehl, M.; Chiaverini, J.; Mehta, K. K.; Romero, J.; Lingaraju, N. B.; Weiner, A. M.; Peace, D.; Cernansky, R.; Lobino, M.; Diamanti, E.; Vidarte, L. T.; Camacho, R. M. 2022 Roadmap on integrated quantum photonics. J. Phys. Photonics 2022, 4 (1), 12501, DOI:10.1088/2515-7647/ac1ef4.

2. Scheller, D.; Hrunski, F.; Schwarberg, J. H.; Knolle, W.; Soykal, Ö. O.; Udvarhelyi, P.; Narang, P.; Weber, H. B.; Hollendonner, M.; Nagy, R. Quantum-enhanced electric field mapping within semiconductor devices. Phys. Rev. Applied 2025, DOI:10.1103/pv13-vgcw.

3. Parthasarathy, S. K.; Kallinger, B.; Kaiser, F.; Berwian, P.; Dasari, D. B. R.; Friedrich, J.; Nagy, R. Scalable quantum memory nodes using nuclear spins in Silicon Carbide. Phys. Rev. Applied 2023, 19 (3), 11048, DOI:10.1103/PhysRevApplied.19.034026.

4. Son, N. T.; Anderson, C. P.; Bourassa, A.; Miao, K. C.; Babin, C.; Widmann, M.; Niethammer, M.; Ul Hassan, J.; Morioka, N.; Ivanov, I. G.; Kaiser, F.; Wrachtrup, J.; Awschalom, D. D. Developing silicon carbide for quantum spintronics. Appl. Phys. Lett. 2020, 116 (19), 190501, DOI:10.1063/5.0004454.

5. Zhou, Y.; Tan, J.; Hu, H.; Hua, S.; Jiang, C.; Liang, B.; Bao, T.; Nie, X.; Xiao, S.; Lu, D.; Wang, J.; Song, Q. Silicon carbide: A promising platform for scalable quantum networks. Applied Physics Reviews 2025, 12 (3), DOI:10.1063/5.0262377.

6. Nagy, R.; Niethammer, M.; Widmann, M.; Chen, Y.-C.; Udvarhelyi, P.; Bonato, C.; Hassan, J. U.; Karhu, R.; Ivanov, I. G.; Son, N. T.; Maze, J. R.; Ohshima, T.; Soykal, Ö. O.; Gali, Á.; Lee,

S.-Y.; Kaiser, F.; Wrachtrup, J. High-fidelity spin and optical control of single silicon-vacancy centres in silicon carbide. Nature communications 2019, 10 (1), 1954, DOI:10.1038/s41467-019-09873-9.

7. Anderson, C. P.; Bourassa, A.; Miao, K. C.; Wolfowicz, G.; Mintun, P. J.; Crook, A. L.; Abe, H.; Ul Hassan, J.; Son, N. T.; Ohshima, T.; Awschalom, D. D. Electrical and optical control of single spins integrated in scalable semiconductor devices. Science 2019, 366 (6470), 1225–1230, DOI:10.1126/science.aax9406.

8. May, A.; Rommel, M.; Baier, L.; Schraml, M.; Dick, J. F.; Jank, M. P. M.; Schulze, J. A 4H-SiC CMOS Technology enabling Smart Sensor Integration and Circuit Operation above 500 °C. 2024 Smart Systems Integration Conference and Exhibition (SSI), Hamburg, Germany, 2024, 1–5, DOI:10.1109/SSI63222.2024.10740550.

9. Wang, H.; Lai, P.; Islam, M. Z.; Hasan, A. S. M. K.; Di Mauro, A.; Anika, N.-E.-A.; Russell, R.; Feng, Z.; Chen, K.; Faruque, A.; White, T.; Chen, Z.; Mantooth, H. A. A review of silicon carbide CMOS technology for harsh environments. Materials Science in Semiconductor Processing 2024, 178, 108422, DOI:10.1016/j.mssp.2024.108422.

10. Yi, A.; Wang, C.; Zhou, L.; Zhu, Y.; Zhang, S.; You, T.; Zhang, J.; Ou, X. Silicon carbide for integrated photonics. Applied Physics Reviews 2022, 9 (3), DOI:10.1063/5.0079649.

11. Krumrein, M.; Nold, R.; Davidson-Marquis, F.; Bouamra, A.; Niechziol, L.; Steidl, T.; Peng, R.; Körber, J.; Stöhr, R.; Gross, N.; Smet, J. H.; Ul-Hassan, J.; Udvarhelyi, P.; Gali, A.; Kaiser, F.; Wrachtrup, J. Precise Characterization of a Waveguide Fiber Interface in Silicon Carbide. ACS Photonics 2024, 11 (6), 2160–2170, DOI:10.1021/acsphotonics.4c00538.

12. Babin, C.; Stöhr, R.; Morioka, N.; Linkewitz, T.; Steidl, T.; Wörnle, R.; Liu, D.; Hesselmeier, E.; Vorobyov, V.; Denisenko, A.; Hentschel, M.; Gobert, C.; Berwian, P.; Astakhov, G. V.; Knolle, W.; Majety, S.; Saha, P.; Radulaski, M.; Tien Son, N.; Ul-Hassan, J.; Kaiser, F.; Wrachtrup, J. Fabrication and nanophotonic waveguide integration of silicon carbide colour centres with preserved spin-optical coherence. Nature materials 2022, 21 (1), 67–73, DOI:10.1038/s41563-021-01148-3.

13. Heiler, J.; Körber, J.; Hesselmeier, E.; Kuna, P.; Stöhr, R.; Fuchs, P.; Ghezellou, M.; Ul-Hassan, J.; Knolle, W.; Becher, C.; Kaiser, F.; Wrachtrup, J. Spectral stability of V2-centres in sub-micron 4H-SiC membranes. npj Quantum Mater 2024, 9 (34), DOI:10.1038/s41535-024-00644-4.

14. Degen, C. L.; Reinhard, F.; Cappellaro, P. Quantum sensing. Rev. Mod. Phys. 2017, 89 (3), 3141, DOI:10.1103/RevModPhys.89.035002.

15. Levine, E. V.; Turner, M. J.; Kehayias, P.; Hart, C. A.; Langellier, N.; Trubko, R.; Glenn, D. R.; Fu, R. R.; Walsworth, R. L. Principles and techniques of the quantum diamond microscope. Nanophotonics 2019, 8 (11), 1945–1973, DOI:10.1515/nanoph-2019-0209.

16. Castelletto, S.; Boretti, A. Silicon carbide color centers for quantum applications. J. Phys. Photonics 2020, 2 (2), 22001, DOI:10.1088/2515-7647/ab77a2.

17. Fuchs, F.; Stender, B.; Trupke, M.; Simin, D.; Pflaum, J.; Dyakonov, V.; Astakhov, G. V. Engineering near-infrared single-photon emitters with optically active spins in ultrapure silicon carbide. Nature communications 2015, 6, 7578, DOI:10.1038/ncomms8578.

18. Hessenauer, J.; Körber, J.; Ghezellou, M.; Ul-Hassan, J.; Astakhov, G. V.; Knolle, W.; Wrachtrup, J.; Hunger, D. Cavity enhancement of V2 centers in 4H-SiC with a fiber-based Fabry–Perot microcavity. Optica Quantum 2025, 3 (2), 175, DOI:10.1364/OPTICAQ.557206.

19. Myers, B. A.; Das, A.; Dartiailh, M. C.; Ohno, K.; Awschalom, D. D.; Bleszynski Jayich, A. C. Probing surface noise with depth-calibrated spins in diamond. Physical review letters 2014, 113 (2), 27602, DOI:10.1103/PhysRevLett.113.027602.

20. Sangtawesin, S.; Dwyer, B. L.; Srinivasan, S.; Allred, J. J.; Rodgers, L. V. H.; Greve, K. de; Stacey, A.; Dontschuk, N.; O'Donnell, K. M.; Di Hu; Evans, D. A.; Jaye, C.; Fischer, D. A.; Markham, M. L.; Twitchen, D. J.; Park, H.; Lukin, M. D.; Leon, N. P. de. Origins of Diamond Surface Noise Probed by Correlating Single-Spin Measurements with Surface Spectroscopy. Phys. Rev. X 2019, 9 (3), DOI:10.1103/PhysRevX.9.031052.

21. Hayashi, K.; Matsuzaki, Y.; Ashida, T.; Onoda, S.; Abe, H.; Ohshima, T.; Hatano, M.; Taniguchi, T.; Morishita, H.; Fujiwara, M.; Mizuochi, N. Experimental and Theoretical Analysis of Noise Strength and Environmental Correlation Time for Ensembles of Nitrogen-Vacancy Centers in Diamond. J. Phys. Soc. Jpn. 2020, 89 (5), 54708, DOI:10.7566/JPSJ.89.054708.

22. Weber, J. R.; Koehl, W. F.; Varley, J. B.; Janotti, A.; Buckley, B. B.; van de Walle, C. G.; Awschalom, D. D. Quantum computing with defects. Proceedings of the National Academy of Sciences of the United States of America 2010, 107 (19), 8513–8518, DOI:10.1073/pnas.1003052107.

23. Preskill, J. Fault-tolerant quantum computation. 1997, 9712048. arXiv quant-ph. https://doi.org/10.48550/arXiv.quant-ph/9712048 (accessed 03 24, 2026).

24. Radulaski, M.; Widmann, M.; Niethammer, M.; Zhang, J. L.; Lee, S.-Y.; Rendler, T.; Lagoudakis, K. G.; Son, N. T.; Janzén, E.; Ohshima, T.; Wrachtrup, J.; Vučković, J. Scalable Quantum Photonics with Single Color Centers in Silicon Carbide. Nano letters 2017, 17 (3), 1782–1786, DOI:10.1021/acs.nanolett.6b05102.

25. Lohrmann, A.; Castelletto, S.; Klein, J. R.; Ohshima, T.; Bosi, M.; Negri, M.; Lau, D. W. M.; Gibson, B. C.; Prawer, S.; McCallum, J. C.; Johnson, B. C. Activation and control of visible single defects in 4H-, 6H-, and 3C-SiC by oxidation. Applied Physics Letters 2016, 108 (2), DOI:10.1063/1.4939906.

26. Abe, Y.; Umeda, T.; Okamoto, M.; Kosugi, R.; Harada, S.; Haruyama, M.; Kada, W.; Hanaizumi, O.; Onoda, S.; Ohshima, T. Single photon sources in 4H-SiC metal-oxide-semiconductor field-effect transistors. Applied Physics Letters 2018, 112 (3), DOI:10.1063/1.4994241.

27. Johnson, B. C.; Woerle, J.; Haasmann, D.; Lew, C.-K.; Parker, R. A.; Knowles, H.; Pingault, B.; Atature, M.; Gali, A.; Dimitrijev, S.; Camarda, M.; McCallum, J. C. Optically Active Defects at the SiC/SiO2 Interface. Phys. Rev. Applied 2019, 12 (4), DOI:10.1103/PhysRevApplied.12.044024.

28. Hijikata, Y.; Komori, S.; Otojima, S.; Matsushita, Y.; Ohshima, T. Impact of formation process on the radiation properties of single-photon sources generated on SiC crystal surfaces. Applied Physics Letters 2021, 118 (20), DOI:10.1063/5.0048772.

29. Nakanuma, T.; Tahara, K.; Kutsuki, K.; Shimura, T.; Watanabe, H.; Kobayashi, T. Control on the density and optical properties of color centers at SiO2/SiC interfaces by oxidation and annealing. Applied Physics Letters 2023, 123 (10), DOI:10.1063/5.0166745.

30. Morioka, N.; Babin, C.; Nagy, R.; Gediz, I.; Hesselmeier, E.; Di Liu; Joliffe, M.; Niethammer, M.; Dasari, D.; Vorobyov, V.; Kolesov, R.; Stöhr, R.; Ul-Hassan, J.; Son, N. T.; Ohshima, T.; Udvarhelyi, P.; Thiering, G.; Gali, A.; Wrachtrup, J.; Kaiser, F. Spin-controlled generation of indistinguishable and distinguishable photons from silicon vacancy centres in silicon carbide. Nature communications 2020, 11 (1), 2516, DOI:10.1038/s41467-020-16330-5.

31. Schwarberg, J. H.; Karhu, R.; Kallinger, B.; Rommel, M.; Schmidt, R.; Schulze, J. Investigation of CMOS Single Process Steps on 4H-SiC a-Plane Wafers for Quantum Applications. 2024 47th MIPRO ICT and Electronics Convention (MIPRO) 2024, 1566–1572, DOI:10.1109/MIPRO60963.2024.10569589.

32. Wang, J.-F.; Li, Q.; Yan, F.-F.; Liu, H.; Guo, G.-P.; Zhang, W.-P.; Zhou, X.; Guo, L.-P.; Lin, Z.-H.; Cui, J.-M.; Xu, X.-Y.; Xu, J.-S.; Li, C.-F.; Guo, G.-C. On-Demand Generation of Single Silicon Vacancy Defects in Silicon Carbide. ACS Photonics 2019, 6 (7), 1736–1743, DOI:10.1021/acsphotonics.9b00451.

33. Choy, K. L., Ed. Chemical vapour deposition (CVD). Advances, technology, and applications, First edition; Materials science and engineering 22; CRC Press: Boca Raton, London, New York, 2019.

34. Song, Y.; Dhar, S.; Feldman, L. C.; Chung, G.; Williams, J. R. Modified Deal Grove model for the thermal oxidation of silicon carbide. Journal of Applied Physics 2004, 95 (9), 4953–4957, DOI:10.1063/1.1690097.

35. Goto, D.; Hijikata, Y. Unified theory of silicon carbide oxidation based on the Si and C emission model. J. Phys. D: Appl. Phys. 2016, 49 (22), 225103, DOI:10.1088/0022-3727/49/22/225103.

36. Kobayashi, T.; Okuda, T.; Tachiki, K.; Ito, K.; Matsushita, Y.; Kimoto, T. Design and formation of SiC (0001)/SiO 2 interfaces via Si deposition followed by low-temperature oxidation and high-temperature nitridation. Appl. Phys. Express 2020, 13 (9), 91003, DOI:10.35848/1882-0786/ababed.

37. Yoshioka, H.; Nakamura, T.; Kimoto, T. Generation of very fast states by nitridation of the SiO2/SiC interface. Journal of Applied Physics 2012, 112 (2), DOI:10.1063/1.4740068.

38. Watanabe, H.; Kobayashi, T.; Iwamoto, H.; Nakanuma, T.; Hirai, H.; Sometani, M. Comprehensive research on nitrided SiO 2 /SiC interfaces by high-temperature nitric oxide annealing formed on basal and non-basal planes. Jpn. J. Appl. Phys. 2025, 64 (1), 10801, DOI:10.35848/1347-4065/ada03c.

39. Guan, J.; Gale, G.; Bersuker, G.; Jeon, Y.; Nguyen, B.; Barnett, J.; Jackson, J. Properties of Chemical Oxides from Pre-Gate Clean Processes and Their Role in the Electrical Thickness of Thermally Grown Ultrathin Gate Oxides. In Cleaning technology in semiconductor device manufacturing: Proceedings of the sixth international symposium; The Electrochemical Society, Ed.; Proceedings volume / Electrochemical Society 99-36; Electrochemical Society: Pennington, NJ, 2000; p77.

40. Franssila, S.; Sainiemi, L. Reactive Ion Etching (RIE). In Encyclopedia of Microfluidics and Nanofluidics: Reactive Ion Etching (RIE); Li, D., Ed.; Springer New York: New York, NY, s.l., 2015; pp 2911–2921.

41. Steidl, T.; Kuna, P.; Hesselmeier-Hüttmann, E.; Di Liu; Stöhr, R.; Knolle, W.; Ghezellou, M.; Ul-Hassan, J.; Schober, M.; Bockstedte, M.; Gali, A.; Vorobyov, V.; Wrachtrup, J. Single V2 defect in 4H Silicon Carbide Schottky diode at low temperature. Nature communications 2025, 16, 4669, DOI:10.1038/s41467-025-59647-9.

42. Jakschik, S.; Schroeder, U.; Hecht, T.; Gutsche, M.; Seidl, H.; Bartha, J. W. Crystallization behavior of thin ALD-Al2O3 films. Thin Solid Films 2003, 425 (1-2), 216–220, DOI:10.1016/S0040-6090(02)01262-2.

43. Triyoso, D.; Liu, R.; Roan, D.; Ramon, M.; Edwards, N. V.; Gregory, R.; Werho, D.; Kulik, J.; Tam, G.; Irwin, E.; Wang, X.-D.; La, L. B.; Hobbs, C.; Garcia, R.; Baker, J.; White, B. E.; Tobin, P. Impact of Deposition and Annealing Temperature on Material and Electrical Characteristics of ALD HfO2. J. Electrochem. Soc. 2004, 151 (10), F220, DOI:10.1149/1.1784821.

44. Jeong, S.-W.; Lee, H. J.; Kim, K. S.; You, M. T.; Roh, Y.; Noguchi, T.; Xianyu, W.; Jung, J. Effects of annealing temperature on the characteristics of ALD-deposited HfO2 in MIM capacitors. Thin Solid Films 2006, 515 (2), 526–530, DOI:10.1016/j.tsf.2005.12.288.

45. Modreanu, M.; Sancho-Parramon, J.; Durand, O.; Servet, B.; Stchakovsky, M.; Eypert, C.; Naudin, C.; Knowles, A.; Bridou, F.; Ravet, M.-F. Investigation of thermal annealing effects on microstructural and optical properties of HfO2 thin films. Applied Surface Science 2006, 253 (1), 328–334, DOI:10.1016/j.apsusc.2006.06.005.

46. Rafí, J. M.; Zabala, M.; Beldarrain, O.; Campabadal, F. Deposition Temperature and Thermal Annealing Effects on the Electrical Characteristics of Atomic Layer Deposited Al2O3 Films on Silicon. J. Electrochem. Soc. 2011, 158 (5), G108, DOI:10.1149/1.3559458.

47. Godignon, P.; Torregrosa, F.; Zekentes, K. Silicon Carbide Doping by Ion Implantation. Materials Research Foundations 2020, 69, 107–174, DOI:10.21741/9781644900673-3.

48. Hallén, A.; Linnarsson, M. Ion implantation technology for silicon carbide. Surface and Coatings Technology 2016, 306, 190–193, DOI:10.1016/j.surfcoat.2016.05.075.

49. Grimaldi, M. G.; Calcagno, L.; Musumeci, P.; Frangis, N.; van Landuyt, J. Amorphization and defect recombination in ion implanted silicon carbide. Journal of Applied Physics 1997, 81 (11), 7181–7185, DOI:10.1063/1.365317.

50. Zhou, Y.; Wang, Z.; Rasmita, A.; Kim, S.; Berhane, A.; Bodrog, Z.; Adamo, G.; Gali, A.; Aharonovich, I.; Gao, W.-B. Room temperature solid-state quantum emitters in the telecom range. Science advances 2018, 4 (3), eaar3580, DOI:10.1126/sciadv.aar3580.

51. Widmann, M.; Lee, S.-Y.; Rendler, T.; Son, N. T.; Fedder, H.; Paik, S.; Yang, L.-P.; Zhao, N.; Yang, S.; Booker, I.; Denisenko, A.; Jamali, M.; Momenzadeh, S. A.; Gerhardt, I.; Ohshima, T.; Gali, A.; Janzén, E.; Wrachtrup, J. Coherent control of single spins in silicon carbide at room temperature. Nature materials 2015, 14 (2), 164–168, DOI:10.1038/nmat4145.

52. Wang, J.; Zhou, Y.; Wang, Z.; Rasmita, A.; Yang, J.; Li, X.; Bardeleben, H. J. von; Gao, W. Bright room temperature single photon source at telecom range in cubic silicon carbide. Nature communications 2018, 9 (1), 4106, DOI:10.1038/s41467-018-06605-3.

53. Schwarberg, J.; Gobert, C.; Hrunski, F.; May, A.; Knolle, W.; Schmid, F.; Rommel, M.; Schulze, J. Scalable Fabrication and Electrical Characterization of Lateral pin-Diodes on 4H-SiC a-Plane Wafers for Functionalization of Silicon Vacancies. Proceedings of Silicon Carbide and Related Materials 2025, in press.

# Supporting Information: Impact of Surface Treatment on Noise in PL-Measurements of Silicon Vacancies in 4H-SiC Lateral pin-Diodes

## Supporting Information Section 1 – Sample Processing Information

For the passivation experiments, a low n-type doped epitaxial layer was grown on a 4H-SiC c-plane wafer. The wafer was diced into 5 mm × 5 mm chips. To minimize the influence of potential variations in the epitaxial layer, all measurements were conducted on adjacent chips.

Prior to processing, all samples were subjected to a standardized cleaning sequence consisting of $HNO_3$, Caro's acid, HF, and TMAH. Immediately before the processes listed in Table SI 1, a time-coupled HF dip was performed to remove the native oxide layer.

Following processing, the thickness of the relevant surface layers was measured using a Horiba Auto SE spectroscopic ellipsometer, employing a spot size of 100 µm × 100 µm.

**Table SI 1.** Overview of the key processing parameters for all samples from the surface modification study grouped by process type.

| Name | Temp. (°C) | Time (mm:ss) | Pressure (mbar) | Gases/ Acids | Type: Power (W) | Thickness (nm) |
|---|---|---|---|---|---|---|
| Plasma Enhanced Deposition (PECVD & PEALD) | | | | | | |
| CCP Oxide | 300 | 1:00 | 1.2 | $SiH_4$, $N_2O$ | RF: 30 | 30 |
| ICP Oxide | 250 | 4:16 | 0.33 | $SiH_4$, $O_2$ | ICP: 350 | 55 |

| Si Nitride | 250 | 2:38 | 0.37 | $SiH_4$, $NH_3$ | ICP: 250 | 61 |
|---|---|---|---|---|---|---|
| Al Oxide | 350 | 500[1] | 20 | TMA, $O_2$ | RF: 400 | 52 |
| Hf Oxide | 350 | 500[1] | 53 | TDMAHf, $O_2$ | RF: 300 | 61 |
| Low Pressure Deposition (LPCVD) | | | | | | |
| Si Oxide | 710 | 12:00 | 0.133 | TEOS | - | 53 |
| Si Nitride | 780 | 2:30 | 0.27 | $SiH_4$, $NH_3$ | - | 51 |
| Thermally Grown Oxides | | | | | | |
| SiC | 1300 | 33:00 | 880 | $O_2$ | - | 59 |
| SiC + NO[2] | 1300 | 60:00 | 880 | NO | - | 59 |
| Poly[3] | 850 | 360:00 | 1013 | $O_2$, DCE | - | 65 |
| Poly + NO[2] | 1300 | 60:00 | 880 | NO | - | 69 |
| Surface Modifications | | | | | | |
| Dry Etch[4] | 20 | 2:30 | 13 | $O_2$, $SF_6$ | ICP: 1500<br>RF: 100 | - |
| ALE[5] | 20 | 200[1] | 0.01 | $Cl_2$, Ar | ICP: 200 | - |
| Resist Asher | 20 | 30:00 | 0.8 | $O_2$ | RF: 800 | - |
| Caro Etching | 140 | 15:00 | - | $H_2O_2$, $H_2SO_4$ | - | < 2 |
| HF-Dip | 20 | 1:00 | - | HF | - | - |

1) For ALD depositions/ ALE etching the number of cycles instead of process time is given.

2) Only the annealing conditions are given in this row. Prior to the annealing the layer growth stated in the row above was carried out.

3) Prior to thermal oxidation a 30 nm thick LPCVD polycrystalline silicon layer was deposited.

4) In addition to the RF power, a bias voltage of 120 V was applied towards the sample to enable directional etching.

5) In addition to the ICP power, a bias voltage of 24 V was applied towards the sample to enable the removal of the modified surface.

Supporting Information Section 2 – PL Measurement Setup

Photoluminescence (PL) characterization was performed using a commercially available confocal PL setup developed by SQUTEC. Excitation was provided by a 730 nm laser operated at a power of 0.25 mW. This laser power was in previous testing found to be a good compromise for low noise signals while still being high enough to get ~8 kcounts/s from a single $V_{Si}$ in a planar sample without any emission enhancing structures. The laser was focused onto the sample using a 100x Zeiss objective with a numerical aperture of 0.9 and a working distance of 1 mm.

To suppress back-reflected excitation light, the emitted PL signal was spectrally filtered using a dichroic mirror with a nominal cutoff at 875 nm, in combination with an 800 nm long-pass filter. The filtered light was coupled into an optical fiber and directed either to a superconducting nanowire single-photon detector (SNSPD), read out using a Swabian Instruments Time Tagger 20, or to an Ocean Optics QEPro-XR300 spectrometer for spectral analysis.

The xy-scans and depth scans were conducted with a pixel scan rate of 30 Hz and a pixel spacing of roughly 100 nm. Due to the high refractive index of SiC (~ 2.7), depth scans into the material appear optically contracted by a factor of ~ 2.7 when referenced to the stage displacement. This geometric contraction arises from refraction at the sample surface, as described by Snell's law. However, since a high-NA objective is used, the simple paraxial approximation slightly overestimates the actual focal depth due to angle-dependent effects such as spherical aberration and limited angular acceptance. A more accurate description requires models such as the Gibson & Lanni model, which account for high-angle refraction and wavefront distortion.[1]

The second-order correlation $g^{(2)}$ measurements were conducted using a fiber-integrated 50/50 beam splitter connected to two separate SNSPD channels, with a temporal bin width of 500 ps.

For cryogenic spectral measurements, the sample was cooled to 4 K using an AttoDry800 closed-cycle cryostat.

## Supporting Information Section 3 – Comparison of PL on c-plane and a-plane wafers

Especially for applications with silicon vacancies 4H-SiC a-plane wafers are of great interest. In these substrates the dipole of the $V_{Si}$ is aligned with the surface[2] enabling resonant excitation from the surface of the wafer. For this study most experiments were carried out on conventional c-plane wafers due to availability. Since the transferability of the results onto a-plane substrates is of great interest, some measurements were also carried out on a-plane wafers. As can be seen in Figure SI 1, the optical properties of the epitaxial layer itself are comparable to those of c-plane wafers. The interface in this measurement shows slightly more bright spots and thus a higher median brightness as can be seen in Figures SI 1b. Together with the results presented in Figure 4 in the main text, which also stem from a-plane samples, transferability of the examination of the optical influence of single process steps (main text – Figure 2) from c-plane to a-plane can be stated.

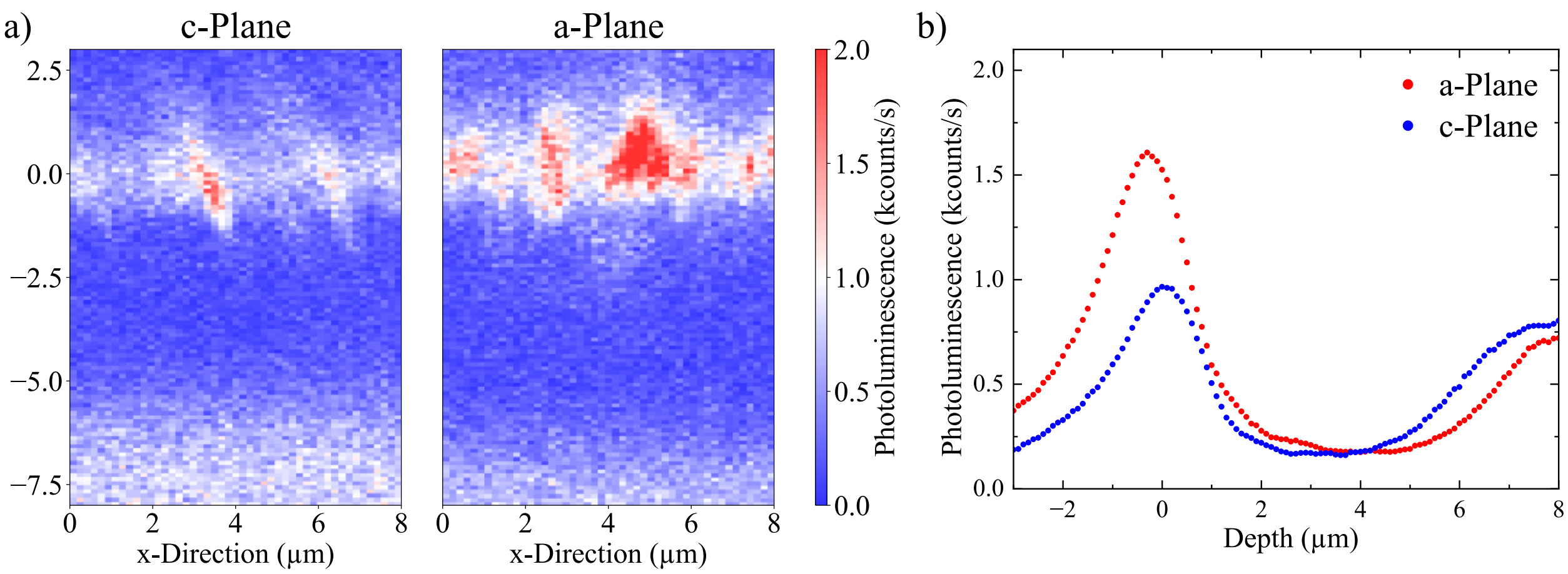

Figure SI 1. Comparison of the PL signal of the epitaxial layers grown on c-plane and a-plane substrates: a) 2D depth scans and b) corresponding PL distribution in dependence of the depth.

Supporting Information Section 4 – Comparison of RIE and ALE surfaces

The accelerated ions from RIE processes are shown to damage the SiC surface. As can be seen in Figure SI 2 (left) this damage consists of a bright background as well as even brighter point defects scattered across the surface. ALE processes use surface modifications with subsequent removal of single atomic layers. As can be seen in Figure SI 2 (middle) no bright defects are generated in this process. Moreover, the damage generated in RIE processes can be reduced by subsequent ALE etching, as shown in Figure SI 2 (right). No difference between the sole ALE and the combination of RIE & ALE can be observed, demonstrating complete removal of the induced damage (Note that the slight difference in brightness for ALE and RIE & ALE is only due to minor deviations of the focal point during scanning).

Nevertheless, the absolute PL of an ALE etched surface is still higher than from the epitaxial layer itself. Since the ALE process works with single atomic layers, this increased PL should stem only from the topmost atomic layers which should be mostly SiC, naturally grown $SiO_2$ or some residuals from the etching process (eg. Cl). Wet chemical etching (like TMAH, HF or Caro`s acid) or the growth of a thin oxide by thermal oxidation, which modify these topmost layers, might be a suitable approach to further improve these surfaces. This might be very beneficial for waveguide or membrane fabrication processes, where dry etching is inevitable.

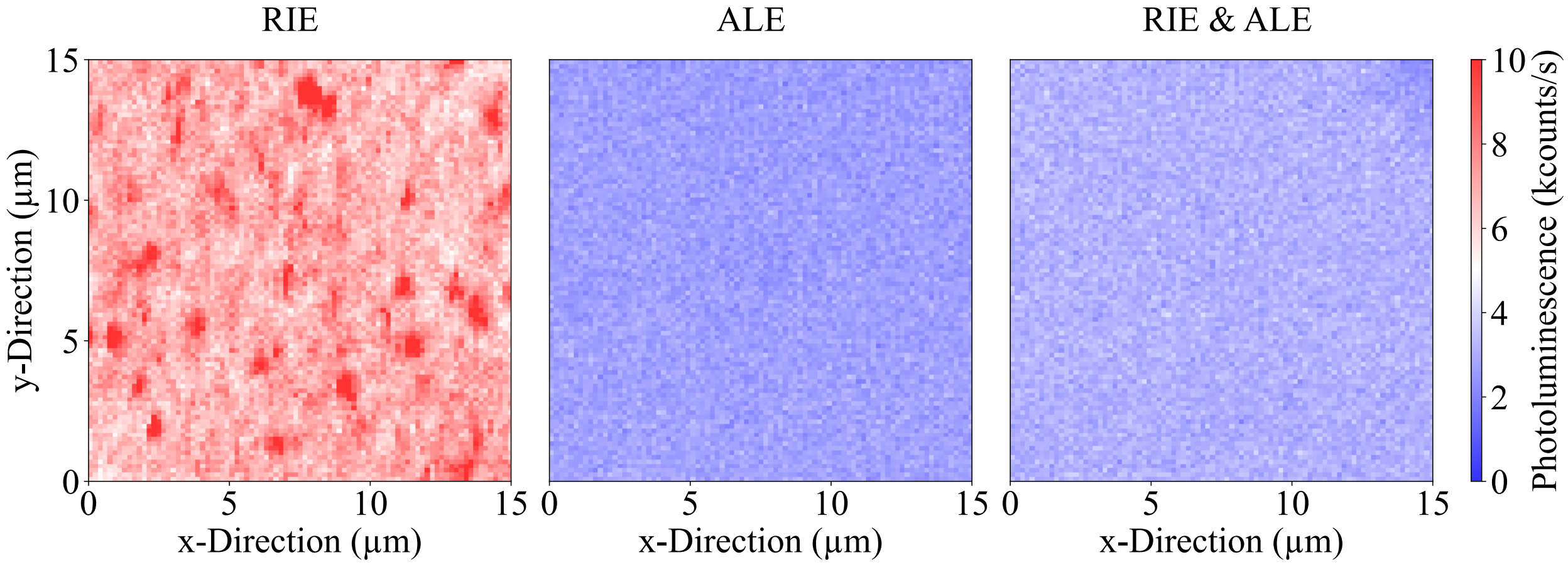

Figure SI 2. Comparison of three xy-Scans of the differently dry etched samples.

## Supporting Information Section 5 – In-line Surface Inspection Tool

The custom-built Intego automated surface inspection system comprises three measurement stations: a bright-field/dark-field (BF/DF) station, a differential interference contrast (DIC) station, and a photoluminescence (PL) station, enabling a broad range of wafer inspection capabilities[3]. For quantum applications, the PL station is of particular relevance. To facilitate the detection of $V_{Si}$ related signals, the system employs a 730 nm excitation LED in combination with an 830 nm long-pass filter. Excitation light is directed onto a ~5 mm × 5 mm region of the wafer. The emitted PL signal passes through the 830 nm long-pass filter to suppress reflected excitation light and is subsequently captured by a camera, yielding a spatially resolved PL image.

To compensate for spatial variations in excitation and collection efficiency, a de-shading correction is applied to the raw data. For full-wafer scans, the wafer is translated beneath the stationary excitation region, and individual images are stitched together to form a composite map. Due to slight mispositioning of the wafer for consecutive measurements slight stitching errors can occur, when subtracting data of two consecutive measurements from each other. Such artifacts are visible in Figure 3b and 3d.

## Supporting Information Section 6 – Etching of the Optical Window

The objective of the etching process is to retain a high-quality thermally grown silicon dioxide layer, which serves both as a low-noise surface passivation and as an electrical passivation layer. To achieve minimal surface damage and low roughness, a highly selective wet chemical etching process is employed. Tetramethylammonium hydroxide (TMAH) is used due to its high selectivity in etching silicon while leaving silicon dioxide unaffected[4].

After the thermally grown oxide is fabricated[5], a sacrificial 500 nm thick polycrystalline silicon (poly-Si) film is deposited via LPCVD to protect the underlying thin oxide. This poly-Si layer also functions as the gate material in the CMOS process[6], thus requiring no additional deposition steps.

Subsequent CMOS-compatible processing involves the deposition of various insulating layers, including oxides and nitrides, for ohmic contact isolation, metal interlayer insulation, and device encapsulation. These layers are structured by high-aspect-ratio dry etching, using the poly-Si layer as a sacrificial buffer to accommodate etch inhomogeneities and necessary overetching. After complete CMOS processing and multiple dry etching steps, the remaining thickness of the poly-Si layer is reduced to below 350 nm, highlighting the necessity of the buffer layer for maintaining oxide integrity.

The residual polycrystalline silicon is removed in a two-step etching process. First, a 240 s dip in 1% HF eliminates the native oxide of the poly-Si layer. This is followed by a self-limiting wet etch using 5% TMAH at 70 °C. Due to the high selectivity of the etch process (~9000:1[4]), precise timing control is not required. After poly-Si removal, the remaining thickness of the thermally grown oxide was measured to be 50 nm (initial thickness: 51 nm), confirming the effectiveness and selectivity of the process.

This etching method is independent of the crystallographic orientation and has been successfully applied to both c-plane and a-plane 4H-SiC wafers.

REFERENCES

1. Gibson, S. F.; Lanni, F. Diffraction by a circular aperture as a model for three-dimensional optical microscopy. Journal of the Optical Society of America. A, Optics and image science 1989, 6 (9), 1357–1367, DOI:10.1364/josaa.6.001357.

2. Nagy, R.; Niethammer, M.; Widmann, M.; Chen, Y.-C.; Udvarhelyi, P.; Bonato, C.; Hassan, J. U.; Karhu, R.; Ivanov, I. G.; Son, N. T.; Maze, J. R.; Ohshima, T.; Soykal, Ö. O.; Gali, Á.; Lee, S.-Y.; Kaiser, F.; Wrachtrup, J. High-fidelity spin and optical control of single silicon-vacancy centres in silicon carbide. Nature communications 2019, 10 (1), 1954, DOI:10.1038/s41467-019-09873-9.

3. Kallinger, Birgit, Schlichting, Holger, Kocher, M, Rommel, Mathias, Berwian, Patrick. Doping-Related Photoluminescence Spectroscopy in 4H-SiC: 13th European Conference on Silicon Carbide and Related Materials, 2020.

4. Wet-chemical etching of silicon and SiO2. Microchemicals - Application Notes. https://www.microchemicals.com/dokumente/application_notes/silicon_etching.pdf.

5. Schwarberg, J. H.; Karhu, R.; Kallinger, B.; Rommel, M.; Schmidt, R.; Schulze, J. Investigation of CMOS Single Process Steps on 4H-SiC a-Plane Wafers for Quantum Applications. 2024 47th MIPRO ICT and Electronics Convention (MIPRO) 2024, 1566–1572, DOI:10.1109/MIPRO60963.2024.10569589.

6. May, A.; Rommel, M.; Baier, L.; Schraml, M.; Dick, J. F.; Jank, M. P. M.; Schulze, J. A 4H-SiC CMOS Technology enabling Smart Sensor Integration and Circuit Operation above 500 °C. 2024 Smart Systems Integration Conference and Exhibition (SSI), Hamburg, Germany, 2024, 1–5, DOI:10.1109/SSI63222.2024.10740550.